\documentclass{PoS}
\usepackage{amsmath,amssymb,amsfonts}
\newcommand{\I}{\ensuremath{\mathrm{i}}}

\newcommand{\tr}{\ensuremath{\mathrm{tr}}}
\newcommand{\str}{\ensuremath{\mathrm{str}}}
\newcommand{\arxiv}[1]{arXiv:\,\href{http://arxiv.org/abs/#1}{{\tt #1}}}
\newcommand{\aetap}{\text{a--}\eta'}
\newcommand{\api}{\text{a--}\pi}

\title{Partially quenched chiral perturbation theory for $\mathcal{N}=1$
supersymmetric Yang-Mills theory}

\ShortTitle{PQChPT for $\mathcal{N}=1$ supersymmetric Yang-Mills theory}

\author{\speaker{Gernot M\"unster}, Hendrik St\"uwe\\
Universit\"at M\"unster, Institut f\"ur Theoretische Physik,\\
Wilhelm-Klemm-Str.~9, D-48149 M\"unster, Germany\\
E-mail: \email{munsteg@uni-muenster.de}}

\abstract{Adding a gluino mass term to $\mathcal{N}=1$ supersymmetric
Yang-Mills theory breaks supersymmetry softly. In order to approach the
supersymmetric continuum limit in numerical simulations with the Wilson
action, the bare gluino mass has to be tuned to the limit of vanishing
renormalised gluino mass. This can be done efficiently by means of the mass
of the adjoint pion, which is, however, an unphysical particle. We discuss
how the adjoint pion can be defined in the framework of partially quenched
chiral perturbation theory. A relation between its mass and the mass of the
gluino, analogous to the Gell-Mann-Oakes-Renner relation of QCD, can be
derived.}

\FullConference{The 32nd International Symposium on Lattice Field Theory\\
		 23-28 June, 2014\\
		 Columbia University New York, NY}
\begin{document}
\section{The Model}
\subsection{$\mathcal{N} = 1$ SUSY Yang-Mills Theory}
The $\mathcal{N} = 1$ supersymmetric Yang-Mills theory (SYM) is the simplest
supersymmetric field theory containing non-Abelian gauge fields. Its
structural complexity is, however, comparable to that of QCD. Its field
content is given by a vector supermultiplet, consisting of a gauge field
$A^{a}_{\mu} (x)$, $a = 1, \dots , N^{2}_{c} - 1$, describing gluons
belonging to the gauge group SU($N_{c}$), a fermionic spinor field
$\lambda^{a} (x)$, obeying the Majorana condition $\bar{\lambda} =
\lambda^{T} C$, and an auxiliary field. The Majorana field describes
gluinos, the superpartners of gluons, and it transforms under the adjoint
representation of the gauge group: ${\mathcal{D}}_{\mu} \lambda^{a} =
\partial_{\mu} \lambda^{a} + g\,f_{abc} A^{b}_{\mu} \lambda^{c}$. The
on-shell Lagrangean of SYM in Euclidean space-time reads
\begin{equation}
{\mathcal{L}} =
\frac{1}{4} F_{\mu\nu}^{\,\,\,a} F_{\mu\nu}^{\,\,\,a} +
\frac{1}{2}\,\bar{\lambda}^{a} \gamma_{\mu} ({\mathcal{D}}_{\mu}
\lambda)^{a}.
\end{equation}
It is invariant under the SUSY transformations
\begin{equation}
\delta A_{\mu}^{a} = 
-2\,\I\,\bar{\lambda}^{a} \gamma_{\mu} \varepsilon,
\quad
\delta \lambda^{a} =
- \sigma_{\mu\nu} F_{\mu\nu}^{\,\,\,a} \varepsilon.
\end{equation}
Being part of the supersymmetrically extended Standard Model, SYM represents
an interesting field theory. It has some similarities to QCD, the important
differences being that gluinos are Majorana particles, and that they are in
the adjoint representation, in contrast to quarks.

The Lagrangean can be extended to include a gluino mass term
$m_{\tilde{g}}\, \bar{\lambda}^{a} \lambda^{a}$. The gluino mass breaks
SUSY softly. The action is only invariant under supersymmetry
transformations in the limit $m_{\tilde{g}} = 0$.

\subsection{Motivation}
SYM is an interesting laboratory for the study of the properties of
supersymmetric models. As in the case of QCD, SYM is characterised by a
number of non-perturbative aspects, which can be investigated in a
lattice-discretised version:
\begin{itemize}
\item
SYM has a discrete chiral symmetry $Z_{2 N_{c}}$, which is predicted to be
broken spontaneously down to $Z_{2}$. The breaking is associated with a
gluino condensate $<\lambda\lambda> \not= 0$.
\item
SYM is expected to show confinement, and the particle states should be bound
states, forming supermultiplets.
\item
Static quarks, belonging to the fundamental representation of the gauge
group, are expected to be confined.
\item
Spontaneous breaking of SUSY is predicted not to occur for SYM.
\item
SUSY is broken by the lattice regularisation. A question which is still open
is if there is a continuum limit in which SUSY is restored?
\item
Predictions about the low-lying particle spectrum from effective Lagrangeans
\cite{Veneziano:1982ah,Farrar:1997fn} should be checked on the lattice.
\end{itemize}

\subsection{SUSY on the Lattice}
Lattice discretisation generically breaks SUSY \cite{Bergner:2009vg}. In the
case of SYM a fine-tuning of the bare gluino mass parameter in the continuum
limit is enough to approach both the (spontaneously broken) chiral symmetry
and supersymmetry of the continuum theory \cite{Curci:1986sm,Suzuki:2012pc}.
Based on this, Curci and Veneziano \cite{Curci:1986sm} proposed to use the
Wilson action and to search for a supersymmetric continuum limit by an
appropriate tuning of parameters. The Wilson action for SYM is given by
\begin{multline}
S = -\frac{\beta}{N_{c}} \sum\limits_{p} \textrm{Re\,Tr}\ U_{p}\\
+ \frac{1}{2}
\sum\limits_{x} \Biggl\{\bar{\lambda}^{a}_{x} \lambda_{x}^{a} -
{\kappa} \sum\limits_{\mu = 1}^{4} \Bigl[ \bar{\lambda}^{a}_{x +
\hat{\mu}} V_{ab, x \mu} (1 + \gamma_{\mu}) \lambda^{b}_{x}
+ \bar{\lambda}^{a}_{x} V^{t}_{ab, x \mu} (1 - \gamma_{\mu})
\lambda^{b}_{x + \hat{\mu}}\Bigr]\Biggr\},
\end{multline}
where $V_{ab, x \mu}$ are the link variables in the adjoint representation.
The parameters in the lattice action are the inverse gauge coupling $\beta =
2 N_{c}/g^{2}$ and the hopping parameter $\kappa = 1/(2 m_{0} + 8)$, related
to the bare gluino mass $m_{0}$.

Numerical simulations of this model, with gauge group SU(2), have been
performed by the M\"unster-DESY-Frankfurt group in recent years, see the
contributions to this conference by P.~Giudice and S.~Piemonte, and
Refs.~\cite{Bergner:2013nwa,Bergner:2013jia}.

\section{The Goals}
\subsection{Phase transition for SU(2)}
As a function of the hopping parameter $\kappa$ the gluino condensate makes
a jump at a certain value $\kappa_{c}$. In the phase diagram the line
$\kappa=\kappa_{c}(\beta)$ represents a first order phase transition.
\begin{figure}[t]
\begin{center}
\includegraphics[width=0.58\textwidth]{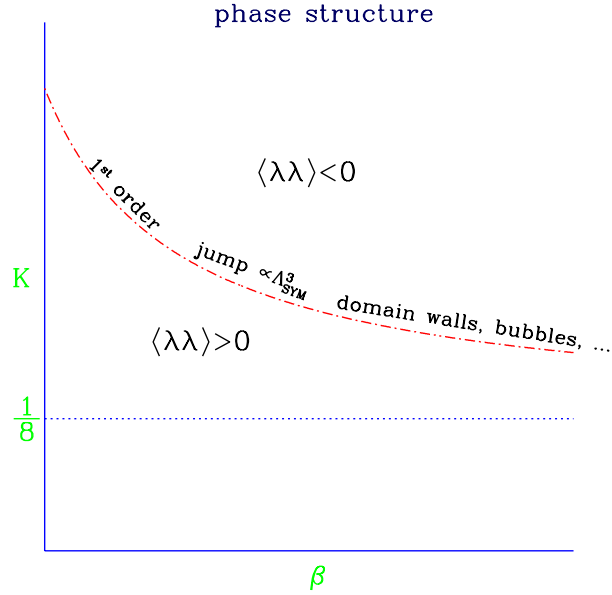}
\caption{The phase diagram of SYM with gauge group SU(2)}
\end{center}
\end{figure}

The recovery of both supersymmetry and chiral symmetry in the continuum
limit requires to tune the hopping parameter to the point $\kappa_c(\beta)$,
where the renormalised gluino mass vanishes
\cite{Curci:1986sm,Suzuki:2012pc}.

\subsection{The adjoint pion}
The gluino mass term is not protected against additive renormalisation in
the Curci-Veneziano formulation. Therefore the point of vanishing gluino
mass is not given a priori, but has to be determined with suitable
observables. A numerically relatively cheap and therefore attractive way to
tune to $\kappa_c$ is to search for the point where the adjoint pion mass
vanishes: $m_{\api} \rightarrow 0$.

However, SYM does not have a continuous chiral symmetry and thus the
spontaneous chiral symmetry breaking is not accompanied by
(pseudo-)Goldstone bosons like pions, whose masses would vanish in the
chiral limit. So, what is the adjoint pion $\api$\,?

A pseudoscalar mesonic bound state, called $\aetap$, is represented by the
interpolating field $\bar{\lambda} \gamma_{5} \lambda$. Its correlator
has connected and disconnected pieces:
\begin{center}
\includegraphics[width=20mm]{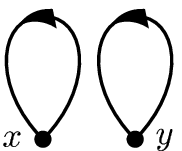}
\quad\raisebox{6mm}{\Large -\ 2\ }
\includegraphics[width=20mm]{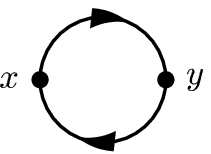}
\end{center}
The correlator of the $\api$ is now given by the connected part of the
$\aetap$ correlator, and the adjoint pion mass can be obtained unambiguously
from it. The $\api$ correlator has the form of the correlator of a meson
formed out of different gluino species. But since SYM only contains one
gluino, the $\api$ is not a physical particle in the Hilbert space of the
theory.

The assumption underlying the tuning of $\kappa$ is that the adjoint pion
mass vanishes with the renormalised gluino mass as
\begin{equation}
m_{\api}^2 \propto m_{\tilde g}\,,
\end{equation}
analogously to the Gell-Mann-Oakes-Renner (GOR) relation of QCD \cite{GOR},
\begin{equation}
m_{\pi}^2 \propto m_{q}\,.
\end{equation}
An argument for this relation, based on the OZI-approximation of SYM, has
been given in \cite{Veneziano:1982ah}.

\begin{figure}[b]
\begin{center}
\includegraphics[angle=270,width=0.9\textwidth]{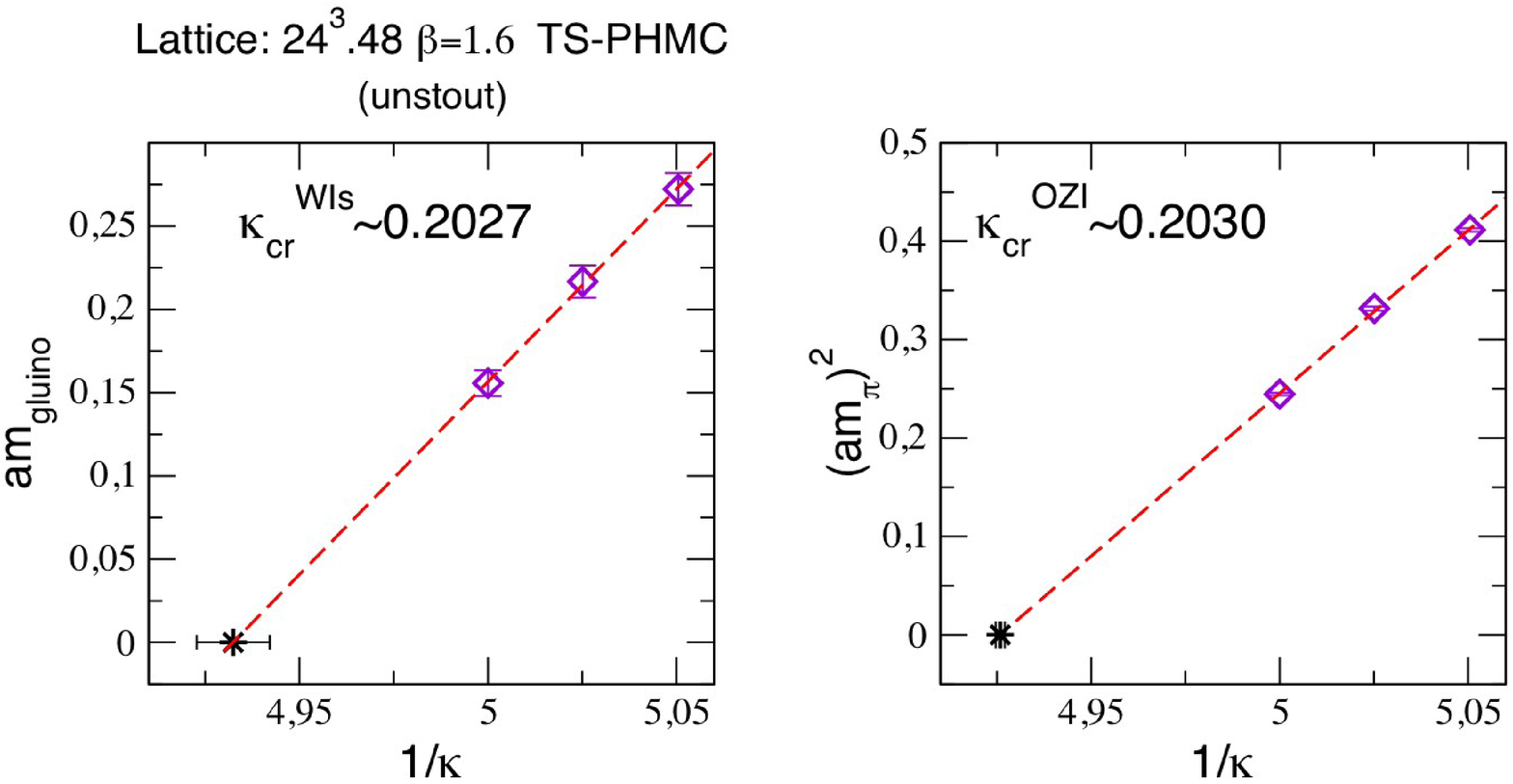}
\caption{The renormalised gluino mass from SUSY Ward identities (left
figure) and the adjoint pion mass squared (right figure) as functions of
$\kappa$}
\end{center}
\end{figure}

On the other hand, the renormalised gluino mass $m_{\tilde g}$ can be
determined by means of the lattice supersymmetric Ward identities as
discussed in \cite{Farchioni:2001wx}. Numerical investigations of both
$m_{\tilde g}$ from supersymmetric Ward identities and $m_{\api}$ have been
performed in ~\cite{Demmouche:2010sf}.
The results are in agreement with
\begin{equation}
a m_{\tilde g} Z_S = \frac{1}{2}
\left( \frac{1}{\kappa} - \frac{1}{\kappa_c} \right),
\qquad
(a m_{\api})^2 \simeq A
\left( \frac{1}{\kappa} - \frac{1}{\kappa_c} \right),
\end{equation}
and support the validity of the above assumption, see Fig.~2. The $\api$,
however, yields a more precise signal for the tuning than the supersymmetric
Ward identities.

\subsection{Goals}
The goals of this work are:
\begin{itemize}
\item
define the adjoint pion $\api$ properly,
\item
establish the relation $m_{\api}^2 \propto m_{\tilde g}$.
\end{itemize}

\subsection{Approach}
In QCD the GOR relation can be derived in the framework of chiral
perturbation theory. Thus the idea is to use this as a starting point for
SYM, too. In contrast to QCD, however, SYM does not have a continuous chiral
symmetry. Therefore the approach consists in adding additional flavours of
gluinos, $\lambda_i(x)$, $i=2,\dots, N$, which are quenched, in order to
keep SUSY intact. This is a particular case of Partially Quenched Chiral
Perturbation Theory (PQChPT), in the spirit of the case of one-flavour QCD
\cite{Farchioni:2007dw}. With the help of the additional gluinos, adjoint
pions can be formed as $\bar{\lambda}_i \gamma_5 (\tau_{\alpha})_{ij}
\lambda_j$ with $i,j = 1,2$.

\section{The Calculation}
\subsection{Adding gluinos}
Let us start by extending SYM with $N-1$ additional flavours of gluinos
$\lambda_i(x)$, $i=2,\dots, N$. In contrast to QCD, where the chiral
symmetry group of $N$ quarks is given by $\mathrm{SU}(N)_L \otimes
\mathrm{SU}(N)_R$, due to the Majorana condition the chiral symmetry group
of extended SYM turns out to be given by a subgroup isomorphic to SU($N$).
If the gluinos are represented as Weyl fermions, this SU($N$) is the group
of transformations of $N$ Weyl fermions.

Spontaneous breakdown of chiral symmetry, accompanied by non-vanishing
gluino condensates, breaks the group $G = \mathrm{SU}(N)$ down to $H =
\mathrm{SO}(N)$. To be specific, we consider the case $N=2$ in the
following. The Goldstone manifold is then the coset space $G/H =
\mathrm{SU}(2)/\mathrm{SO}(2) \sim S^2$. It can be parameterised by $u =
\exp (\I\alpha_1 T_1 + \I\alpha_3 T_3)$, where $T_i$ are the generators of
SU(2). It is now convenient to define the nonlinear Goldstone boson field by
$U(x) = u(x)^2 = u(x) u(x)^T \doteq \exp \left(\I \,\phi(x)/F\right)$,
because the transformation law of this group valued field,
\begin{equation}
U(x) \rightarrow U'(x) = V U(x) V^T\,, \quad V \in \textrm{SU}(2),
\end{equation}
is similar to that of the usual chiral perturbation theory.

Analogously to the approach used in QCD, the leading order effective
Lagrangean can be determined to be
\begin{equation}
\mathcal{L}_{2} 
= \frac{F^2}{4} \tr ( \partial_{\mu} U \partial^{\mu} U^{\dagger})
+ \frac{F^2}{4} \tr ( \chi U^{\dagger} + U \chi ^{\dagger}),
\end{equation}
where
\begin{equation*}
\chi = 2 B_0 m_{\tilde{g}} \mathbf{1}
\end{equation*}
is the symmetry breaking mass term, and $F$ and $B_0$ are low-energy
constants.

Note that the theory with 2 gluinos might be conformal or near-conformal
\cite{Athenodorou:2013eaa}, implying a different breaking pattern. However,
its discussion here just serves as a preparation of the following PQChPT
analysis, which is not affected by this possibility.

\subsection{PQChPT}
In order that the dynamical content of the model is identical to that of
SYM, and the correlation functions of the original fields are unchanged, the
additional gluinos have to be quenched, which means that they are not taken
into account in the fermionic functional integral. This is a case of PQChPT
\cite{Bernard:1993sv,Sharpe:1997by}. Partial quenching can be described
theoretically by the introduction of bosonic ghost fermions \cite{Morel}, in
our case ghost gluinos. The contribution of the ghost gluinos exactly
cancels the contribution of the additional gluinos, and only the
contribution of the original single gluino remains. In the case of $N=2$
there is a single ghost gluino $\rho(x)$, compensating the contribution of
the additional gluino. The resulting chiral symmetry group is the graded
group SU(2$|$1), and the Goldstone boson field is a graded matrix field
\begin{equation}
\phi = 
\begin{pmatrix}
\phi_{ss} & \phi_{sv} & \phi_{sg} \\
\phi_{vs} & \phi_{vv} & \phi_{vg} \\
\phi_{gs} & \phi_{gv} & \phi_{gg}
\end{pmatrix},
\end{equation}
where $s$, $v$ and $g$ stand for sea, valence and ghost. Now the adjoint
pion can in this formulation be defined to be the meson represented by
$\phi_{sv}$.

The leading order effective Lagrangean for the partially quenched theory is
given by
\begin{equation}
\mathcal{L}_{2}^{PQ} 
= \frac{F^2}{4}\, \str ( \partial_{\mu} U \partial^{\mu} U^{\dagger})
+ \frac{F^2}{4}\, \str ( \chi U^{\dagger} + U \chi ^{\dagger}),
\end{equation}
where $\str$ denotes the supertrace. The next-to-leading order terms can be
constructed analogously to the NLO terms for QCD \cite{GL1}, and are not
reproduced here. They contain further low-energy constants $L_i$, the
so-called Gasser-Leutwyler coefficients.

The masses of the pseudo-Goldstone bosons can be calculated in PQChPT by
means of the effective Lagrangean. We have calculated them in NLO. Whereas
the tree-level contributions are similar to the ones in QCD, the loop
contributions differ due to the different group structure. For the mass of
the adjoint pion $m_{\api}$ we find
\begin{equation}
m^2_{\api} = 2 B_0 m_{\tilde g}
+ \frac{(2 B_0 m_{\tilde g})^2}{F^2} (30 L_8 - 2 L_4 - 7 L_5 + 8 L_6),
\end{equation}
with the low-energy coefficients $L_i$ mentioned above. 
For small $m_{\tilde g}$ we recognise the desired GOR-relation
\begin{equation}
m^2_{\api} = 2 B_0 m_{\tilde g} .
\end{equation}

\subsection{Results}
To summarise, the results of this investigation are:
\begin{itemize}
\item
The adjoint pion $\api$ is defined in PQChPT,
\item
$m^2_{\api} = 2 B_0 \,m_{\tilde{g}}$
in leading order PQChPT.
\end{itemize}
Details can be found in \cite{Munster:2014cja}.

\end{document}